# Field Tunable Magnetic Transitions of CsCo$_2$(MoO$_4$)$_2$(OH): A Triangular Chain Structure with a Frustrated Geometry

Liurukara D Sanjeewa,[a,b*] V. Ovidiu Garlea,[c] Randy S. Fishman,[d] Mahsa Foroughian,[e] Li Yin,[d] Jie Xing,[d] David S. Parker,[d] Tiffany M. Smith Pellizzeri,[f] Athena S. Sefat,[d] Joseph W. Kolis[d*]

[a]University of Missouri Research Reactor (MURR), University of Missouri, Columbia, MO 65211, USA
[b]Department of Chemistry, University of Missouri, Columbia, MO 65211, USA
[c]Neutron Scattering Division, Oak Ridge National Laboratory, Oak Ridge, TN 37831, USA
[d]Materials Science and Technology Division, Oak Ridge National Laboratory, Oak Ridge, TN 37831, USA
[e]Department of Chemistry and Center for Optical Materials Science and Engineering Technologies (COMSET), Clemson University, Clemson, SC 29634-0973, USA
[f]Department of Chemistry and Biochemistry, Eastern Illinois University, Charleston, IL 61920, USA

*Corresponding Authors

**Abstract**

Identifying and characterizing new magnetic systems with Co$^{2+}$ ions can enhance our understanding of quantum behavior since Co$^{2+}$ can host a pseudospin-1/2 magnetic ground state. Understanding the magnetic ground state and the phase diagrams of such systems are central to the development of new theoretical models to described emergent quantum properties of complex magnetic systems. The sawtooth chain compound, CsCo$_2$(MoO$_4$)$_2$(OH), is one such complex magnetic system and here, we present a comprehensive series of magnetic and neutron scattering measurements to determine its magnetic phase diagram. The magnetic properties of CsCo$_2$(MoO$_4$)$_2$(OH) exhibit a strong coupling to the crystal lattice and its magnetic ground state can be easily manipulated by applied magnetic fields. There are two unique Co$^{2+}$ ions, base and vertex, with $J_{bb}$ and $J_{bv}$ magnetic exchange. The magnetism is highly anisotropic with the $b$-axis (chain) along the easy axis and the material orders antiferromagnetically at $T_N$ = 5 K. There are two successive metamagnetic transitions, the first at $H_{c1}$ = 0.2 kOe into a ferrimagnetic structure, and the other at $H_{C2}$ = 20 kOe to a ferromagnetic phase. Heat capacity measurements in various fields support the metamagnetic phase transformations, and the magnetic entropy value is intermediate between $S$ = 3/2 and 1/2 states. The zero field antiferromagnetic phase contains vertex magnetic vectors (Co(1)) aligned parallel to the $b$-axis, while the base vectors (Co(2)) are canted by 34° and aligned in an opposite direction to the vertex vectors. The spins in parallel adjacent chains align in opposite directions, creating an overall antiferromagnetic structure. At a 3 kOe applied magnetic field, adjacent chains flip by 180° to generate a ferrimagnetic phase. An increase in field gradually induces the Co(1) moment to rotate along the $b$-axis and align in the same direction with Co(2) generating a ferromagnetic structure. The antiferromagnetic exchange parameters are calculated to be $J_{bb}$ = 0.028 meV and $J_{bv}$ = 0.13 meV, while the interchain exchange parameter is considerably weaker at $J_{ch}$ = (0.0047/$N_{ch}$) meV. Our results demonstrate that the CsCo$_2$(MoO$_4$)$_2$(OH) is a promising candidate to study new physics associated with sawtooth chain magnetism and it encourages further theoretical studies as well as the synthesis of other sawtooth chain structures with different magnetic ions.

1. **Introduction**

Geometrically frustrated magnetic materials with non-trivial magnetic states at low temperatures can provide an exciting playground to study exotic phenomena in condensed matter physics. [1-5] The quantum spin liquid (QSL) state is one such exotic quantum state where conventional Neel order is not reached due to the large system degeneracy. Even in the disordered state (liquid-like state) at near $T$ = 0 K, the magnetic spins are strongly interacting on an energy scale of an ordered magnet or even higher. [6-13] The realization of novel QSL materials and associated phenomena could be the key to develop next generation quantum communication and computation.

The triangular magnetic lattice is a canonical geometry in theories of QSLs, wherein spins are entangled in a long-range fluctuating ground state. Moreover, triangular lattices (TL) are considered to be the simplest model of a geometrically frustrated magnetic framework, with isolated magnetic ions forming equilateral triangles via linkages with nonmagnetic bridging groups like halides or oxyanions. However, these triangular lattices could be distorted due to the structural perturbations like lattice distortion or the presence of unequal crystallographic sites that are magnetic. In addition to structural distortions, anisotropy and next nearest neighbor interactions could also lift the degeneracy of the magnetic ground sate. The splitting of the magnetic ground state in degenerate systems is very interesting since it can produce complex magnetic phase diagrams with multiple field induced magnetic transitions. Therefore, these complex triangular materials could offer routes to identify new magnetic ground states and open up avenues to develop new theoretical tools. [14-19]

Non-trivial magnetic states driven by magnetic frustration have also been observed in one-dimensional (1-D) systems, particularly in triangular chains. The delta chain (also known as "sawtooth chains") and half-delta chain represent special classes of 1-D compounds that possess geometrical frustration. [20-22] The quantum nature and spin frustration of such 1-D systems could manifest themselves in the behavior of the elementary excitations. [23-31] Structurally speaking, the sawtooth magnetic lattice is made by sharing the corners of triangles alternately pointing up and down in a wave fashion along a 1-D chain. Figure 1 depicts the more relevant magnetic interactions of both half-delta chain and the delta chain. If the magnetic triangles of the half-delta chain and the delta chain are isosceles, the two nearest-neighbor magnetic exchange interactions can be identified as $J_1$ and $J_2$. Here, $J_1$ is the base-apex ($J_{bv}$) interaction and $J_2$ is the base-base ($J_{bb}$) interactions. If the $J_1$ and $J_2$ are equal within the triangle ($J_1 = J_2 = J > 0$), the magnetic ground state of a spin-1/2 delta chain is a spin-singlet. In this scenario, the lowest excitation in the sawtooth chain can be described by forming *kink* and *antikink* pairs within the periodic boundary conditions. This special ground state has dispersionless elementary excitations with a gap of $\Delta E = 0.234J$. [23-31] In general, only a handful of sawtooth chain compounds have been synthesized and characterized so far. Furthermore, experimental realization of sawtooth lattice structures with frustrated magnetism is even more scarce. The few examples are delafossite-$YCuO_{2.5}$, [32-33] euchroite-$Cu_2(AsO_4)(OH)$ $3H_2O$ [34] and olivines-$ZnLn_2S_4$ ($Ln$ = Er and Yb)-type structures. [35] These compounds do not show any long-range magnetic ordering down to milli-K temperatures, which means they could be promising QSLs candidates. On the other hand, distortion of the sawtooth magnetic lattices could generate different interactions ($J_1 \neq J_2$). In such a scenario the magnetic ground state is determined by the relative strength of $J_1$ and $J_2$, and could lead to exotic magnetically ordered states. [20-22] Furthermore, the sawtooth magnetic chains are often interconnected by nonmagnetic spacers such as oxyanion groups ($SiO_4^{4-}$, $PO_4^{3-}$, $AsO_4^{3-}$, $VO_4^{2-}$, etc.), [20-22] which serve as paths for possible interchain interactions. Such complexity has been observed in $A_2BX_4$ ($A$ = Mn, Fe, Ni; $B$ = Si, Ge; $X$ = S, Se, Te O) olivine structures, [36-40] $Rb_2Fe_2O(AsO_4)_2$, [20] $CuFe_2Ge_2$, [41] $Cu_2Cl(OH)_3$, [24] $Fe_2O(SeO_3)_2$, [42] and $\{[Cu(bpy)(H_2O)][Cu(bpy)(mal)(H_2O)]\}(ClO_4)_2$ (bpy = 2,2'-bipyridine and mal = malonate dianion). [43-44]

For most of these systems, a complete understanding of the magnetic interactions remains challenging due to the unavailability of sizable, high quality single crystals to perform detailed anisotropic magnetic property characterization, especially in regard to external applied magnetic fields. In recent years we undertook a systematic investigation of the complex magnetic properties of oxyanion-based sawtooth lattices. The highly anisotropic $Rb_2Fe_2O(AsO_4)_2$ is one such example which orders antiferromagnetically ($T_N$ = 25 K) and exhibits a complex magnetic phase diagram with field induced sates. [20] We were also successful in synthesizing novel sawtooth magnetic materials using the hydrothermal method. For example, both NaCo$_2$(SeO$_3$)$_2$(OH) and Rb$_2$Mn$_3$(MoO$_4$)$_3$(OH)$_2$ was synthesized hydrothermally. The NaCo$_2$(SeO$_3$)$_2$(OH) structure is a rare experimental realization of the sawtooth magnetic lattice with two competing magnetic transitions that undergoes a FM transition at 11 K and followed by an AFM transition at 3.8 K. Additionally, NaCo$_2$(SeO$_3$)$_2$(OH) exhibits a complex field induced magnetic phase diagram. [22] Recently, Rb$_2$Mn$_3$(MoO$_4$)$_3$(OH)$_2$ was synthesized as single crystals using a high-temperature ($T$ = 580 ˚C) hydrothermal method. The compound has a half-delta chain magnetic lattice and exhibits complex magnetism with two consecutive magnetic transitions which include a transition from a paramagnetic to an incommensurate phase below 4.5 K and to a commensurate antiferromagnetic transition below 3.5 K. [21] This wide variety of magnetic ground states in sawtooth chain compounds motivate us to synthesize new materials, and therefore we have continued to explore the MoO$_4^{2-}$ oxyanion based sawtooth compounds.

Recently we realized a series of novel sawtooth structures, Cs$M$$_2$(MoO$_4$)$_2$(OH), $M$ = Fe, Mn and Co. [45] The MoO$_4^{2-}$ tetrahedra are non-magnetic building blocks that isolate the magnetic lattice by preserving the low-dimensionality of the magnetic structures. They can readily form a variety of extended structures in combination with open shell transition metal octahedra. The MoO$_4^{2-}$ anion could also introduce geometrical frustration since it contains trigonal symmetry and the presence of empty $d$-orbital on Mo$^{6+}$ could facilitate next-nearest neighbor (NNN) interactions between the sawtooth chains (interchain interactions). In some cases, these NNN interactions become stronger and have a direct influence on the magnetic ground state. Considering the few cases of molybdenum based oxyanion transition metal sawtooth chain compounds, we extended our magnetic property characterization to study $CsCo_2(MoO_4)_2(OH)$.

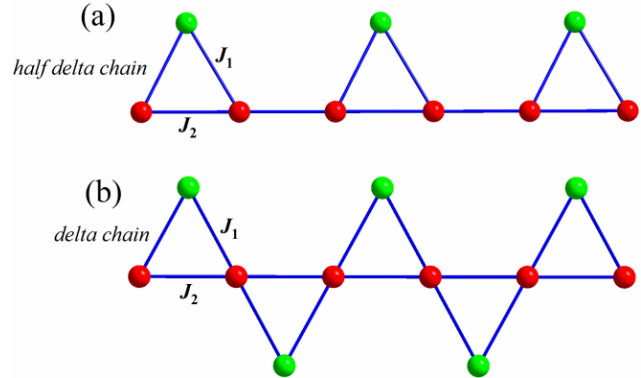

**Figure 1.** Schematic structures of half-delta-chain and delta-chain systems. The $J_1$ and $J_2$ represent the nearest neighbor interactions between the magnetic ions.

In this paper, the phase diagram of sawtooth chain compound, $CsCo_2(MoO_4)_2(OH)$, was studied using multiple experimental methods such as *dc*-magnetic susceptibility, magnetization and specific heat. Large single crystals (~2-3 mm) of $CsCo_2(MoO_4)_2(OH)$ were synthesized using a high-temperature ($T$ = 590 °C) hydrothermal method, which allowed us to perform detailed orientation-dependent magnetic measurements. This compound exhibits a pronounced anisotropic behavior in the magnetic susceptibility and isothermal magnetization, with several field-induced metamagnetic transitions along the Co–O–Co chain direction (*b*-axis). Temperature and field dependent neutron powder diffraction measurements were used to determine the magnetic orders and demonstrate that $CsCo_2(MoO_4)_2(OH)$ exhibits a non-colinear antiferromagnetic ground sate below $T_N$ = 5 K. The magnetic field induces a succession of magnetic phase transformations, to a ferrimagnetic and canted ferromagnetic structure and these are described from the neutron diffraction data.

## 2. Experimental Section

*2.1. Hydrothermal Synthesis of CsCo$_2$(MoO$_4$)$_2$(OH)*

A high temperature ($T$ = 580 °C) high-pressure method was previously employed to synthesize CsCo$_2$(MoO$_4$)$_2$(OH) single crystals. [45] This method allowed us to grow ~2 mm single crystals (Figure 2). However, the yield of this method was very low and resulted in other impurity phases. Therefore, a different approach (low-temperature hydrothermal) was used to grow high yield of single crystals samples of CsCo$_2$(MoO$_4$)$_2$(OH). In a typical reaction, Cobalt(II) acetate (Co(CH$_3$COO)$_2$) and Molybdenum trioxide (MoO$_3$) were used in a stoichiometric ratio of 1 : 1 ratio with 10 mL of 2 M CsCl. Finally, 0.2 mL of acetic acid was added to the solution. The mixture was transferred and sealed in a Teflon-lined stainless-steel autoclave and heated at 200 °C for 3 days. After cooling to room-temperature, pure blue colored microcrystalline sample was recovered using filtration. Several sets of reactions were performed to obtain 3 g of powder sample to perform the neutron powder diffraction experiment. The purity of each batch was checked using powder X-ray diffraction.

*2.2 Magnetic property characterization*

Temperature-dependent and field dependent magnetic measurements were performed using a Quantum Design Magnetic Property Measurement System (MPMS). The single crystal specimen was affixed to a quartz rod using GE varnish and temperature dependent magnetization measurements were carried out along the *b*-axis and perpendicular to the *b*-axis from 2 to 350 K in the applied magnetic field up to 50 kOe. Additionally, anisotropic isothermal magnetization measurements were performed between 2-100 K up to 130 kOe magnetic field using a physical property measurement system (PPMS). The heat capacity ($C_p$) of the sample was measured using PPMS between 2-50 K under 0 and 110 kOe.

*2.3 Neutron scattering*

Neutron powder diffraction measurements were performed on the ground single crystals sample of $CsCo_2(MoO_4)_2(OH)$ using the HB2A Powder Diffractometer at High Flux Isotope Reactor (HFIR) at Oak Ridge National Laboratory. [46] Diffraction patterns were collected using the 2.41 Å wavelength provided by the (113) reflection of a germanium vertical focusing monochromator. Powder samples were pressed into pellets in order to prevent rotation of the powder grains in the applied magnetic field. The pellets were loaded into a vanadium can and inserted in a vertical field superconducting magnet. Data was collected for several magnetic fields up to 45 kOe at 2 K. Additional diffraction measurements were performed using the POWGEN instrument at the Spallation Neutron Source (SNS). [47] For those measurements, the powder samples were loaded into 8 mm diameter vanadium holders and cooled down to 25 K by using a closed-cycle refrigerator. Data were collected with the neutron wavelength band centered at 1.333 Å. The neutron diffraction data were analyzed by using the FullProf Suite Package. [48] The refined NPD pattern at 25 K is depicted in Figure 2a.

**Table 1.** Crystallographic data of $CsCo_2(MoO_4)_2(OH)$ determined by single crystal X-ray diffraction.

| empirical formula | $CsCo_2(MoO_4)_2(OH)$ |
|---|---|
| formula weight (g/mol) | 587.66 |
| crystal system | monoclinic |
| space group, $Z$ | $P2_1/m$ (no.19), 4 |
| Crystal dimensions, mm | 0.08 x 0.02 x 0.02 |
| $T$, K | 298 |
| $a$, Å | 8.2234(3) |
| $b$, Å | 6.0429(2) |
| $c$, Å | 9.0956(3) |
| $\beta$, ° | 99.54(1) |
| volume, Å$^3$ | 445.73(3) |
| $D$(calc), g/cm$^3$ | 4.379 |
| $\mu$ (Mo K$\alpha$), mm$^{-1}$ | 10.452 |
| $F$(000) | 538 |
| $T$max, $T$min | 1.0000, 0.68708 |
| $2\theta$ range | 2.27-30.39 |
| reflections collected | 5412 |
| data/restraints/parameters | 888/1/82 |
| final $R$ [$I > 2\sigma(I)$] $R_1$, $R_{w2}$ | 0.0161, 0.0365 |
| final $R$ (all data) $R_1$, $R_{w2}$ | 0.0239, 0.0397 |
| GoF | 1.006 |
| largest diff. peak/hole, e/ Å$^3$ | 0.0749/-0.560 |

**Table 2.** Selected bond distances (Å) and angles (°) of $CsCo_2(MoO_4)_2(OH)$.

|  | **Co(1)O$_6$** | **Co(2)O$_6$** |  |
|---|---|---|---|
| Co(1)–O(1) | 2.051(4) | Co(2)–O(1) x 2 | 1.993(2) |
| Co(1)–O(2) | 1.989(4) | Co(2–O(2) x 2 | 2.158(3) |
| Co(1)–O(3) x 2 | 2.179(3) | Co(2)–O(3) x 2 | 2.162(3) |
| Co(1)–O(4) x 2 | 2.114(3) |  |  |
| **Mo(1)O$_4$** |  | **Mo(2)O$_4$** |  |
| Mo(1)–O(2) | 1.744(4) | Mo(1)–O(4) x 2 | 1.760(3) |
| Mo(1)–O(3) x 2 | 1.803(3) | Mo(1)–O(5) | 1.815(4) |
| Mo(1)–O(6) | 1.729(4) | Mo(1)–O(7) | 1.733(4) |
|  |  |  |  |
| Co(1)–O(1)–Co(2) | 101.30(1) |  |  |
| Co(1)–O(3)–Co(2) | 92.30(1) |  |  |
| Co(2)–O(1)–Co(2) | 98.55(1) |  |  |
| Co(2)–O(5)–Co(2) | 88.67(1) |  |  |

**Table 3.** Fractional atomic coordinates and isotropic or equivalent isotropic displacement parameters (Å$^2$) of CsCo$_2$(MoO$_4$)$_2$(OH) from single crystal X-ray data measured at room temperature.

| Atom | Wyckoff pos. | x | y | z | Ueq |
|---|---|---|---|---|---|
| Cs(1) | 2e | 0.58807(4) | 0.75000 | 0.70027(5) | 0.01911(3) |
| Co(1) | 2e | 0.11063(9) | 0.75000 | 0.73208(8) | 0.00974(3) |
| Co(2) | 2a | -1.00000 | 0.00000 | 0.50000 | 0.00781(1) |
| Mo(1) | 2e | 0.12857(6) | 0.75000 | 0.34939(5) | 0.00911(2) |
| Mo(2) | 2e | 0.32077(6) | 0.25000 | 0.84493(5) | 0.00948(4) |
| O(1) | 2e | 0.1450(4) | 0.75000 | 0.9607(4) | 0.00912(4) |
| O(2) | 2e | 0.0419(5) | 0.75000 | 0.5118(4) | 0.01533(4) |
| O(3) | 4f | 0.0679(3) | 0.5063(4) | 0.2395(3) | 0.01069(4) |
| O(4) | 4f | 0.2820(3) | 0.4888(5) | 0.7340(3) | 0.01284(4) |
| O(5) | 2e | 0.1837(5) | 0.2500 | 0.9821(4) | 0.01157(1) |
| O(6) | 2e | 0.3411(5) | 0.7500 | 0.3960(5) | 0.02038(4) |
| O(7) | 2e | 0.5250(5) | 0.25000 | 0.9319(5) | 0.03154(4) |
| H(1) | 2e | 0.249(5) | 0.75000 | 1.014(8) | 0.05000(4) |

## 3. Results

### 3.1. Crystal structure of CsCo$_2$(MoO$_4$)$_2$(OH)

The single crystal sample was recovered as blue columns (Figure 2a) with an average size of ~2-3 mm in length. Single crystal X-ray diffraction indicate that CsCo$_2$(MoO$_4$)$_2$(OH) crystallizes in monoclinic crystal structure, space group of $P2/1m$ (No. 19). The unit cell parameters are $a$ = 8.2234(3) Å, $b$ = 6.0429(2) Å, $c$ = 9.0956(3) Å, $\beta$ = 99.54(1)°, $V$ = 445.73(3) Å$^3$. The detail crystallographic data are reported in Tables 1 and 2. Furthermore, our SXRD results confirmed that CsCo$_2$(MoO$_4$)$_2$(OH) is stoichiometric with no detectable site disorder and the crystal structure is shown in Figure 2. The structure contains one crystallographically unique Cs$^+$, two Co$^{2+}$, two Mo$^{4+}$, seven O$^{2-}$ and one H$^+$ ions. Cobalt atoms are found in two different Wyckoff sites, Co(1): 2e (0.11063, 0.75, 0.73208) and Co(2): 2a (-1.0, 0, 0.5) and form CoO$_6$-octahedra (*oct*). Each Mo site (2e-site) is bonded to four oxygen atoms to form MoO$_4$-tetrahedra (*tet*). A partial polyhedral view of the CsCo$_2$(MoO$_4$)$_2$(OH) structure along the *b*-axis is shown in Figure 2b. Triangular chains of $CoO_6$-*oct* are found in a simple packing of *A-A-A* along the *a*-axis. In the topology of a triangular sawtooth chain, $Co(1)O_6$-*oct* forms the vertex of the triangle and two $Co(2)O_6$-*oct* form the base of the triangle, respectively, Figure 2c. Within the each $Co_3$–triangle $Co(1)O_6$-*oct* share edges with two $Co(2)O_6$-*oct* via O(1), O(3) and O(5) which can be best described as $[Co_3O_{13}]$ triangular units with O(1) serving as the $\mu_3$-oxo vertex. The O(1) atom binds to H by forming the only ⁻OH groups in the structure. A Co–Co chain is highlighted by connected bonds in blue in Figure 2c. These chains are bridged by the $MoO_4$-*tet* group along the *c*-axis and form layers of Co–O–Mo–O–Co. These layers are stacked along the *a*-axis. Additionally, triangular Co-delta chains are decorated by the $MoO_4$-*tet* groups forming the top and the bottom of each individual chain, and $Cs^+$ ions fill the gaps between the layers. A summary of Co–O, Mo–O bond distances and Co–O–Co bond angles are given in Table 3. The six coordinate $CoO_6$ units of $CsCo_2(MoO_4)_2(OH)$ are highly distorted and possess an average Co–O distance of 2.104(3) Å occurring over a range from 1.989(4) to 2.179(3) Å, comparable to the expected sum of the Shannon crystal radii, 2.14 Å, for a 6-coordinate high spin $Co^{2+}$ and $O^{2-}$. [49] One may also notice that these highly distorted $CoO_6$-*oct* have smaller bond angles than ideal. For example, O(3)−Co(1)−O(4) and O(1)−Co(2)−O(3) bond angles are 82.08(3)° and 83.95(3)°, respectively, and such small angles are feasible because of the longer Co−O within the $CoO_6$-*oct*: Co(1)−O(3) (2.179(3) Å) in $Co(1)O_6$-*oct* and Co(2)−O(3) (2.162(3) Å) in $Co(2)O_6$-*oct*. These longer bonds propagate between the neighboring $Co(1)O_6$ and $Co(2)O_6$ in a zigzag pattern forming a long range distortion within the individual sawtooth chains. The $MoO_4$-*tet* exhibit a relatively smaller distortion compared to the $CoO_6$-*oct*. The Mo–O bond distances range from 1.729(4) to 1.815(4) Å with an average Mo–O bond distance of 1.761 Å which is consistent with the expected sum of the Shannon crystal radii, 1.79 Å, for a 4-coordinate high spin $Mo^{6+}$ and $O^{2-}$. [49]

Since $Co^{2+}$ is a $3d$ ($d^7$) transition metal, the strength of magnetic exchange coupling is determined by the bond distances and bond angles of $CoO_6$ units. Therefore, it is important to discuss the Co–O bond distances and Co–O–Co bond angles of $CsCo_2(MoO_4)_2(OH)$. In the sawtooth chain, Co(1)–Co(2) and Co(2)–Co(2) distances are 3.1280(4) and 3.0206(1) Å, respectively. Based on these interatomic bond distances we can define unequal interactions occurring between two Co-sites ($J_{bb} \neq J_{bv}$). The bond angles Co(2)–O(1)–Co(2) and Co(2)–O(5)–Co(2) associated with $J_{bb}$ are 88.72(5)° and 98.58(4)°, respectively, while for $J_{bv}$ the angles of Co(1)–O(1)–Co(2) and Co(1)–O(3)–Co(2) are 92.35(5)° and 101.37(5)°, respectively. One may also notice that adjacent sawtooth chains along the $c$-axis can connect via $Mo(1)O_4$ groups. The distance between chains is ~5.2 Å, which can also enable possible interchain interactions, but we would expect much stronger intrachain interaction compared to the interchain interaction via molybdate groups, leading to pseudo-1-D type magnetic properties. Nevertheless, given the structural complexity and competition between $J_{bv}$, $J_{bb}$ and interchain interactions one can expect complex magnetism and a significant anisotropy of $CsCo_2(MoO_4)_2(OH)$.

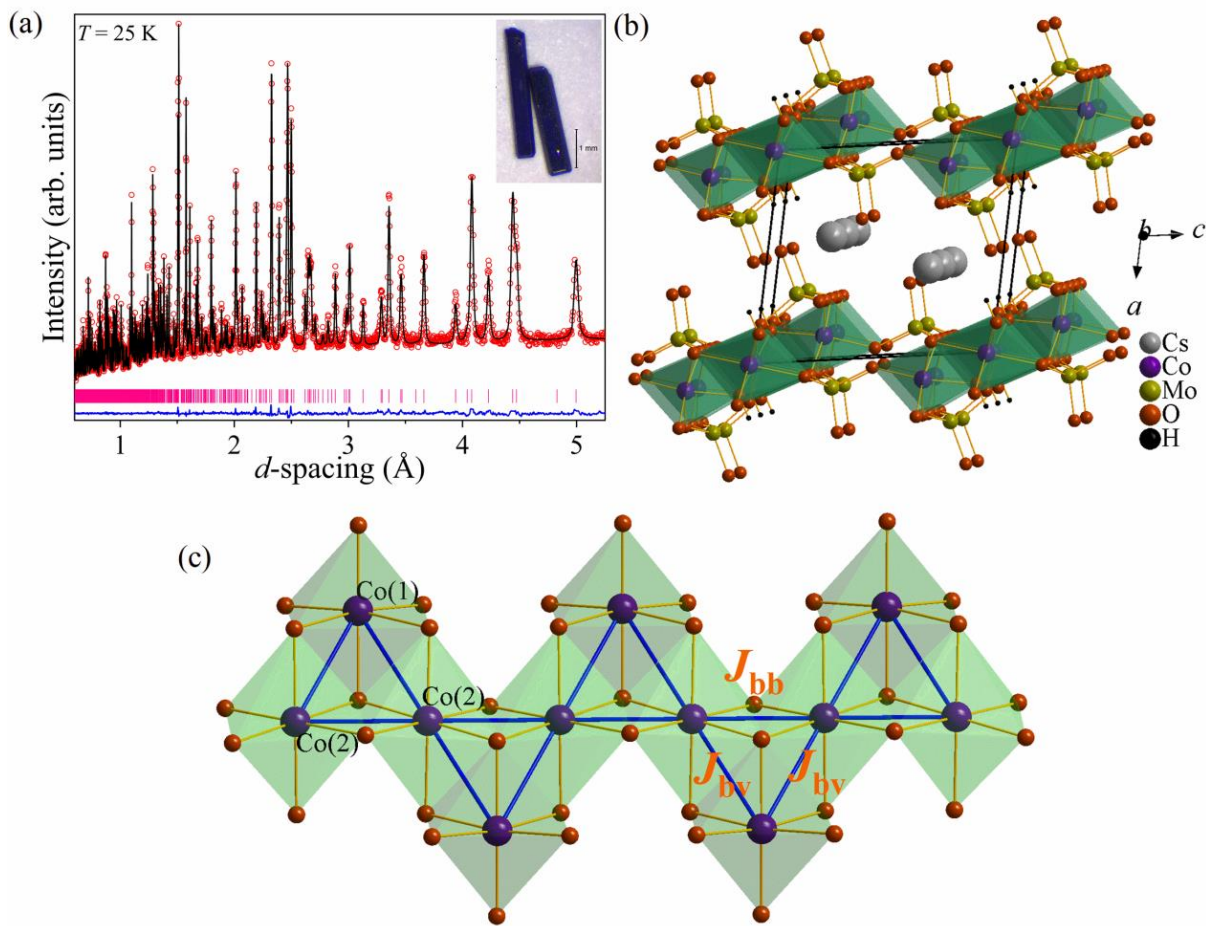

**Figure 2. Rietveld refinement and the crystal structure of $CsCo_2(MoO_4)_2(OH)$.** (a) The best structural model fits using the single crystal structure solution of monoclinic $P2_1/m$ (no.19), symmetry to neutron powder diffraction collected using POWGEN at 25 K. The positions of nuclear Bragg peaks are indicated by pink tick marks. The black line, red circle markers and blue lines represent the model, data and difference curves, respectively. Hydrothermally grown single crystals of $CsCo_2(MoO_4)_2OH$ are displayed in the insert. (b) Partial polyhedral view of $CsCo_2(MoO_4)_2(OH)$ projected along the *b*-axis, showing packing of Co–O–Mo layers on the *ac*-plane. The $Cs^+$ ions reside between the layers. (c) Partial structure of Co–O–Co sawtooth chains made from edged sharing $CoO_6$ octahedra along the *b*-axis. The unequal $J_{bb}$ and $J_{bv}$ exchange interactions are shown using the solid blue line. The Co(1)–Co(2) and Co(2)–Co(2) distances are 3.1280(4) and 3.0206(1) Å, respectively.

### 3.2 Magnetic Properties of CsCo$_2$(MoO$_4$)$_2$(OH)

The temperature dependent magnetization, $\chi = M/H$ of CsCo$_2$(MoO$_4$)$_2$(OH) was measured in both zero-field cooling (ZFC) and field cooling (FC) modes between 2-350 K with a magnetic field parallel and perpendicular to the sawtooth chain (*b*-axis) as shown in Figure 3a. The magnetic susceptibility measured with a low applied field, $H$ = 100 Oe, is shown in Figure 3a. In this low magnetic field, the magnetic susceptibility exhibits a sharp rise below 10 K and produces a peak at $T_N$ = 5 K, suggesting AFM ordering of CsCo$_2$(MoO$_4$)$_2$(OH). The ZFC and FC $M/H$ curves did not bifurcate below 5 K, which excludes spin-glass behavior. At higher temperature above 30 K, magnetic susceptibility measurements along the two crystallographic orientations ($H//b$ and $H\perp b$) fully overlap with each other (paramagnetic region), but a strong anisotropy was observed between the two crystal orientations near the transition temperature, Figure 3a. We found that the inverse magnetic susceptibility in the paramagnetic region for a low applied magnetic field ($H$ = 100 Oe) can be fit using a Curie-Weiss model $M/H = C (T-\theta)$. The data in the paramagnetic region is not strictly linear and also resulted in unreasonable values. Therefore, we performed Curie-Wiess fits to the magnetic susceptibility obtained at a much higher field, $H$ = 10 kOe, Figure 3a inset. Using the fit above 100 K, an effective moment of 5.4 $\mu_B$/Co and a Weiss temperature of -1.5 K were obtained. The negative $\theta_{cw}$ suggests the overall antiferromagnetic nature of CsCo$_2$(MoO$_4$)$_2$(OH). Despite the fact that CsCo$_2$(MoO$_4$)$_2$(OH) has a geometrically frustrated magnetic lattice, the frustration index $f = |\theta_w/T_N|$ is smaller than one. This is different much smaller than in other, similar oxyanion-based sawtooth structures such as Rb$_2$Fe$_2$O(AsO$_4$)$_2$ ($f$ = ~20) [20] and half-sawtooth chain Rb$_2$Mn$_3$(MoO$_4$)$_3$(OH)$_2$ ($f$ = ~24). [21] However, the sawtooth chain structure NaCo$_2$(SeO$_3$)$_2$(OH) also possesses a smaller frustration index, $f$ = ~1.[22] The effective magnetic moment of CsCo$_2$(MoO$_4$)$_2$(OH) is higher than the spin only value of Co$^{2+}$ ($S$ = 3/2, $\mu_{eff}$ = 3.8 $\mu_B$/Co). This could be due to the significant orbital contribution since orbital moments are unquenched for $Co^{2+}$ in an octahedral environment ($t_{2g}^5 e_g^2$, $S = 3/2$, $L = 3$, gs = $^4T_{1g}$). Similar higher values have been noticed previously for pseudo-1-D $Co^{2+}$ compounds such as $SrCo(VO_4)(OH)$ (5.4 $\mu_B$/Co), [50] $Na_4CoTeO_6$ (5.14 $\mu_B$/Co) [51] and $NaCo_2(SeO_3)_2(OH)$ (4.84 $\mu_B$/Co). [22] However, it is important to emphasize that at lower temperatures the ground state of distorted octahedral $Co^{2+}$ is a Kramers doublet. Therefore, at lower temperatures the magnetic properties of $Co^{2+}$ is determined by the lowest Kramers doublet, $J = 1/2$. The effective spin-1/2 ground state at lower temperatures has been reported in many $Co^{2+}$-triangular magnetic materials. [52-53] The presence of such a state in $Co^{2+}$ can be confirmed by measuring the crystal field excitation using high-energy inelastic neutron scattering. [22,54] As an example, our previous work on crystal field excitations of $NaCo_2(SeO_3)_2(OH)$ reveals the existence of a $J_{eff} = 1/2$ state. Additionally, the magnetic structure refinement of $NaCo_2(SeO_3)_2(OH)$ yield an average of 2.1 $\mu_B$ /Co, indicating a possible $J_{eff} = 1/2$ state at lower temperature. [22]

The behavior of the magnetic susceptibility was also measured under multiple fields within the range of 10–50 kOe and it was found that the magnetic behavior depends greatly on the applied magnetic field, indicating field induced magnetic phase transitions at $T < 10$ K (Figure 3b). Magnetic susceptibilities measured at 20, 30 and 50 kOe with the applied magnetic field along the sawtooth chain direction are displayed in the inset of Figure 3b. When the applied magnetic field is parallel to the sawtooth chain direction, a relatively smaller magnetic field ($H = 500$ Oe) induces a magnetic transition from AFM to a FM-like state. This is further confirmed by isothermal magnetization data as we discuss below. The magnetic susceptibility measured at $H = 10$ kOe along the sawtooth chain direction exhibits a rounded maximum at 6.5 K and a further increase of field (up to 50 kOe) pushes the transition to a higher temperature, suggesting a gradual development of a second field-induced transition. On the other hand, the magnetic susceptibility measured for $H \perp b$ exhibits a sharp inflection point at ~5 K up to an applied magnetic field of 1 kOe. This suggests that the first field induce transition corresponds to a spin-flip-type transition along the sawtooth chain to produce a ferro- or ferrimagnetic state in $CsCo_2(MoO_4)_2(OH)$ and is followed by a continuous rotation of the Co-spin along the sawtooth chain to form another FM-like state at higher magnetic field ($H > 0.5$ kOe). This behavior was clearly observed in our isothermal magnetization data and confirmed by using the neutron powder diffraction under applied magnetic field (see below).

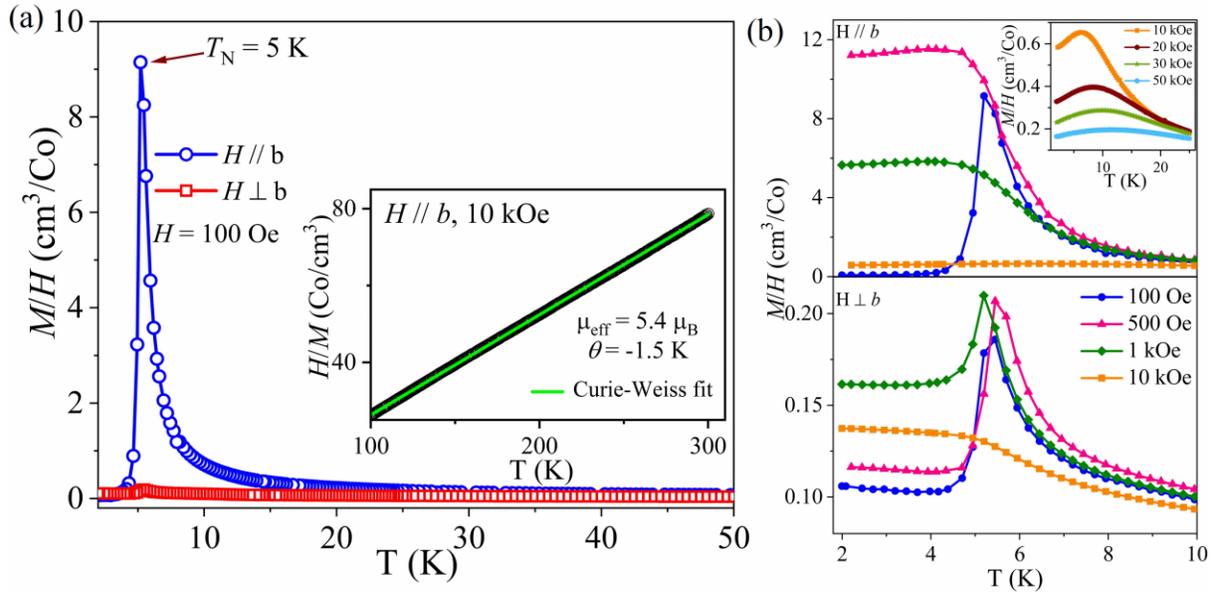

**Figure 3. Anisotropic magnetic susceptibility of $CsCo_2(MoO_4)_2(OH)$.** (a) Magnetic susceptibilities, $\chi = M/H$, of $CsCo_2(MoO_4)_2(OH)$ as a function of temperature under an applied field of 100 Oe along the two crystallographic orientations. Inset: $CsCo_2(MoO_4)_2(OH)$ crystal highlighting the crystallographic $b$-axis (Co–O–Co chain axis) as the longest axis of the crystals. Temperature dependence of the inverse magnetic susceptibility $\chi^{-1}$ in the temperature range 100−300 K. The solid line represents the Curie−Weiss fitting to the data from 100 to 300 K. (b) Field dependent magnetic susceptibility of $CsCo_2(MoO_4)_2(OH)$ measured in different magnetic fields ($H = 100$ Oe – 50 kOe) below 10 K. Results are shown for $H//b$ (top) and $H \perp b$ (bottom).

Figure 4a depicts the isothermal magnetization measured at various temperatures $T = 2, 4, 10, 15, 20$ and 100 K up to 130 kOe. Two clear metamagnetic transitions with magnetization jumps

Page | 16

at critical fields of $H_{c1}$ = 0.2 kOe and at $H_{c2}$ = 20 kOe were observed. These metamagnetic transitions vanish with increasing temperature as shown in Figure 4a. The first metamagnetic transition starts at a relatively smaller field $H_{c1}$ = 0.2 kOe which corresponding to a 1 $\mu_B$/Co jump in the magnetization, Figure 4a. inset. After that, magnetization at 2 K produces a plateau that extends to approximately $H_{c2}$ = 20 kOe, which was assigned as the second critical field. After the broad plateau, magnetization increases monotonously with no saturation, reaching 2.1 $\mu_B$/Co at the highest measured magnetic field of 130 kOe. The first anomaly at $H_{c1}$ could be related to a spin-flip transition within the sawtooth chain and arranged along the *b*-axis (easy axis), or it could be also due to re-arranging of the neighboring sawtooth chains along the *b*-axis by aligning them parallel to each other to generate a net FM moment. Since the second field induced transition happens within a broad range of field, we can assume that the field induced state should evolve gradually as a function of magnetic field, as observed in Figure 3b. On the other hand, isothermal magnetization data perpendicular the Co–O–Co chain direction do not exhibit any field induced transitions even up to 100 kOe, confirming the highly anisotropic nature of this compound. The overall isothermal magnetization data suggests that the system undergoes several field-induced transitions starting from an AF (zero field structure) state to several FM-like states. Stepwise magnetic transitions have been reported in the sawtooth compounds of $Rb_2Fe_2O(AsO_4)_2$ and $Fe_2Se_2O_7$. In $Rb_2Fe_2O(AsO_4)_2$, a metamagnetic transition occurs near 3 kOe at 2 K when the magnetic field is parallel to the *b*-axis (sawtooth chain axis) while on the other hand, $Fe_2Se_2O_7$ features a metamagnetic transition at the relatively higher field $H$ = 55 kOe. None of the reported sawtooth compounds, however, exhibit *successive* field-induced transitions (at least within the measured magnetic fields). [20,42] Therefore, $CsCo_2(MoO_4)_2(OH)$ is a rare structure that exhibits a complex magnetic phase diagram, and an ideal case to study emergent magnetic phenomena associated with the sawtooth chain structures.

To further elucidate the field induced magnetic properties of $CsCo_2(MoO_4)_2(OH)$, the temperature dependent heat capacity was measured up to 110 kOe. Figure 4b shows the heat capacity as a function of temperature of the title compound from 1 to 12 K. We must also note that here the magnetic field was applied perpendicular to the *b*-axis (flat surface) of the single crystal due to the difficulty of aligning the column crystal parallel to the heat capacity platform. At zero applied magnetic field a sharp λ-anomaly peak at 5.1 K was observed which agrees well with our magnetic data (Figure 4b). With increasing magnetic field, the transition temperature remains constant up to 5 kOe, but then it starts to shift to higher temperatures as the magnetic field increases. This indicates that $CsCo_2(MoO_4)_2(OH)$ transformed into a predominantly ferromagnetic phase under a large external magnetic field, which is consistent with the magnetic data, as shown in Figure 3b. To calculate the magnetic entropy, the heat capacity data of $CsCo_2(MoO_4)_2(OH)$ above 15 K were fit by $C_p = \gamma T + \beta_1 T^3 + \beta_1 T^5 + \beta_1 T^7$ where $\gamma = 0$ for an insulator and $\beta_1 T^3 + \beta_1 T^5 + \beta_1 T^7$ is the contribution form the lattice ($C_l$). The magnetic contribution, $C_{mag}$, was calculated as the difference of $C_p$-$C_l$. The magnetic entropy, $S_M$, was estimated by integrating $(C_{mag}/T)dT$. As displayed in the inset of Figure 4b, the magnetic entropy calculated is $S_M = 7.56$ J mol$^{-1}$ K$^{-1}$. The expected entropy ($\Delta S$) for $Co^{2+}$ in $S = 3/2$ spin state is 11.5 J mol$^-$ K$^-$ ($\Delta S = R\ln(2S+1)$), so the observed magnetic entropy is 65% of $R\ln(2S+1)$ and significantly larger than what is expected for an effective spin $S_{eff} = 1/2$, $R\ln2 = 5.76$ J mol$^{-1}$ K$^{-1}$. This result is not uncommon in $Co^{2+}$-based compounds and it can be due to several reasons such as releasing residual entropy below 2 K, more entropy releasing above the transition due to the spin fluctuations, highly anisotropic behavior of the system or a transition from a low-dimensional ordered state to full long-range order in three dimensions. [50, 55-56]

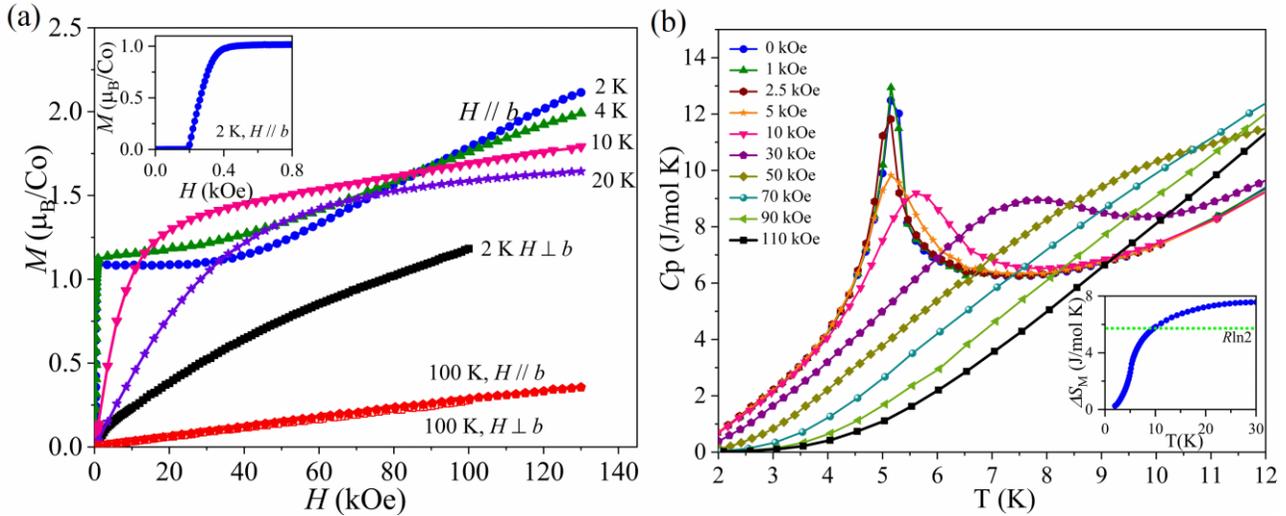

**Figure 4. Isothermal magnetization data and heat capacity of CsCo$_2$(MoO$_4$)$_2$(OH).** (a) Isothermal magnetization data within the temperature range of 2-100 K. The data collected by applying a magnetic field parallel and perpendicular to the *b*-axis. (b) The specific heat data measured under various fields in the 0 – 110 kOe range. The inset shows the magnetic entropy of CsCo$_2$(MoO$_4$)$_2$(OH).

*3.3 Magnetic Structure of CsCo$_2$(MoO$_4$)$_2$(OH) at zero field*

The nuclear structure refinement was performed using the data collected from time-of-flight (TOF) neutron powder diffractometer POWGEN. [47] Data was collected at 25 K using the frame with a center wavelength of 1.33 Å. The obtained crystallographic parameters are consistent with our single crystal X-ray results. The magnetic structure characterization was performed with and without the applied magnetic field using HB-2A powder diffractometer at High Flux Isotope Reactor. [46] The HB-2A measurements were performed using the 2.41 Å wavelength within the range of applied magnetic field of 0 - 4.5 T. Here, a pelletized sample was used to prevent the grain orientation of the powder with the applied magnetic field. Upon cooling the sample below ~ 5 K, additional Bragg peaks were observed, distinct from the nuclear reflections (Figure 5a inset) which indicate the onset of antiferromagnetic long-range magnetic ordering of $CsCo_2(MoO_4)_2(OH)$. The magnetic reflections can be clearly observed by comparing the 30 K and 2 K diffraction patterns, Figure 5a inset. No other extra peaks appear at higher Q below the magnetic transitions and no additional intensity change in the nuclear structure Bragg peaks, which rules out the presence of a structural phase transition with temperature. The magnetic peaks were indexed using the K-Search program included in the FullProf Suite program. [48] These reflections indicate a magnetic propagation vector $k = (0.5, 0, 0)$. Representational analysis was performed using the SARA*h*-program [57] and the possible magnetic space groups were explored using MAXMAGN program at the Bilbao Crystallographic Server. [58] There are four possible irreducible representations (IRs) associated with $k = (0.5, 0, 0)$ and $P2_1/m$ space-group symmetry. The decomposition of the magnetic representations for Co(1) can be written as $\Gamma_{mag} = 1\Gamma^1_1 + 2\Gamma^1_2 + 2\Gamma^1_3 + 1\Gamma^1_4$ and for Co(2) $\Gamma_{mag} = 3\Gamma^1_1 + 0\Gamma^1_2 + 3\Gamma^1_3 + 0\Gamma^1_4$. The best fit to the data was obtained by adopting a magnetic structure model based on the irreducible representation $\Gamma^1$. This model implies a Co(1) magnetic moment that is constrained to be along the *b*-direction, whereas the Co(2) moment can orient along any crystallographic axis. The magnetic space group corresponding to this configuration is $P_a2_1/m$ (#11.55) for a magnetic unit cell: $2a \times b \times c$. The refinement yield an ordered moments of $m_{Co(1)} = 1.3(1)$ $\mu_B$ and $m_{Co(2)} = 3.7(1)$ $\mu_B$ ( $m_a = -1.9(1)$ $\mu_B$, $m_b = -3.10(4)$ $\mu_B$, $m_c = 0.5(2)$ $\mu_B$). The resulted moment for the Co(2) is consistent with the expected moment for $Co^{2+}$ with $S = 3/2$, $\mu_{eff} = 3.8$ $\mu_B$/Co. In contrast, the Co(1) moment is significantly lower compared to the expected moments of $Co^{2+}$ with $S = 3/2$ state. The previously reported sawtooth structures also exhibit a similar behavior, which could be due to the frustrated nature of the magnetic lattice. This difference between the refined magnetic moment and the expected moment may also indicate a possible spin-state transition from $S = 3/2$ to $S = 1/2$ (spin crossover) at lower temperatures. This type of transition is feasible due to the octahedral environment of the $Co^{2+}$ and the presence of the crystal electric field and spin-orbital coupling which lead to a Kramers doublet ground state. [52-53] However, the underlying mechanism for such behavior remains to be determined and further investigations of the crystal electric field and spin Hamiltonian need to be pursued. More work focusing on the synthesis and neutron scattering measurements of $Co^{2+}$-based novel sawtooth structures is of further interest and could help us to understand the different spin states in these compounds.

A representation of the magnetic structure is shown in Figure 5. According to the proposed model, Co(1) moments (green arrows) are aligned parallel to each other within the chain along the *b*-axis, and Co(2) moments (red arrows) arrange in a zigzag pattern and point in opposite direction to the Co(1) moments. The canting angle of the Co(2) moment is about 34°, nearly following one of the Co–O octahedral bonds. With the Co(1) and Co(2) moments pointing in opposite directions, each individual sawtooth chain is characterized by a net ferromagnetic moment along the chain direction (*b*-axis). Within the 2*a*,*b*,*c* magnetic unit cell, these chains arrange antiparallel to each other giving a net AF structure (AF layers on *ab*-plane). We note that this magnetic structure is similar to the structure observed in $Rb_2Fe_2O(AsO_4)_2$ [20] however, it is significantly different from the Co-based sawtooth system $NaCo_2(SeO_3)_2(OH)$. [22]

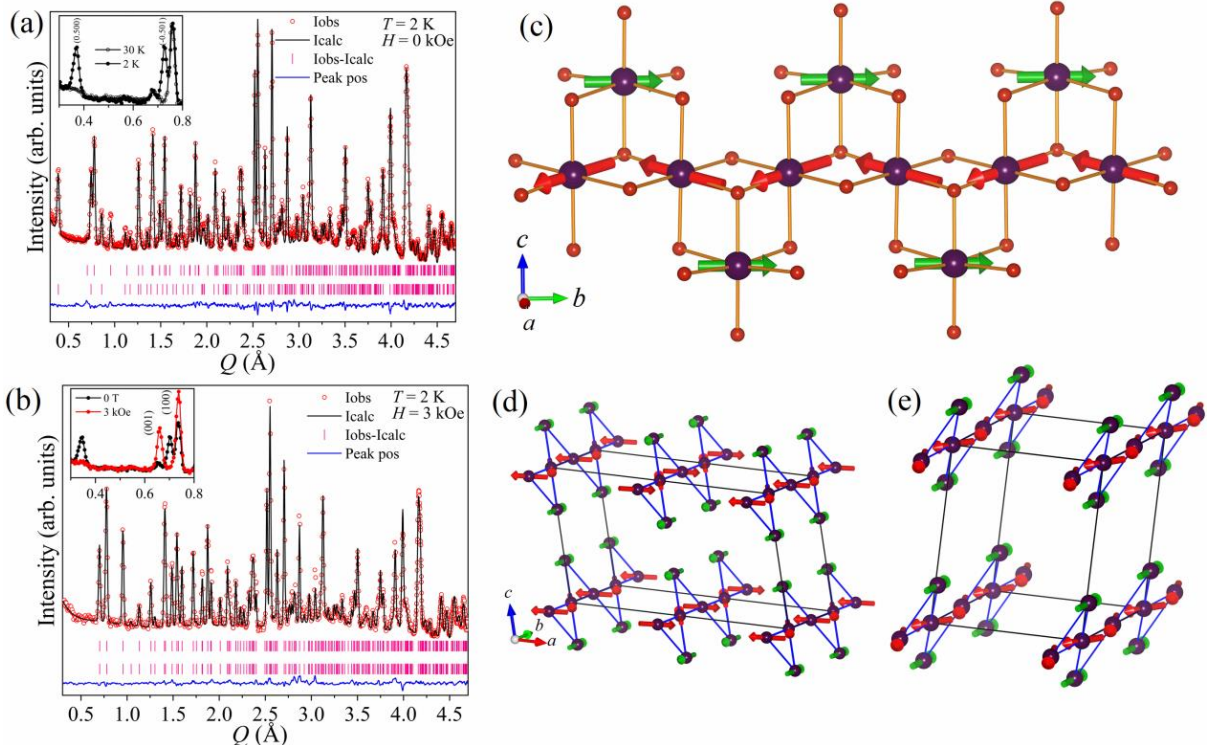

**Figure 5. Magnetic structure refinements.** Rietveld plot of neutron powder diffraction data collected at 2 K in the zero magnetic field (a) and 3 kOe (b). (c) Schematic view of spin structure of Co–O–Co sawtooth chain at 2 K, the green arrows represent the magnetic moments of Co(1) and red arrows represent the Co(2). The nearest neighbor Co distances are depicted as solid blue lines. The Co(1) magnetic moments are collinear with the *b*-axis, while the Co(2) moments are forming undulating patterns, running in opposite direction with respect to the Co(1) moments. (d) The ferrimagnetic sawtooth chains are coupled antiferromagnetically within the *ab*-plane of the unit cell. (e) Magnetic structure at 2 K and applied magnetic field of 3 kOe. In contrast to the zero-field magnetic structure, the polarity of the ferrimagnetic chains remains the same throughout the lattice, yielding an overall ferrimagnetic ordering.

*3.4 Magnetic Structure of CsCo$_2$(MoO$_4$)$_2$(OH) in the applied magnetic field*

Having determined the zero filed magnetic ground state we now turn to investigate the field evaluation of the magnetic ground state of CsCo$_2$(MoO$_4$)$_2$(OH). As described earlier, CsCo$_2$(MoO$_4$)$_2$(OH) exhibit a very sharp metamagnetic transition at 0.2 kOe, Figure 4a. Since the first metamagnetic transition happens in a relatively smaller field, we applied 3 kOe magnetic field which is well above the $H_{c1}$. The magnetic field diminishes the intensity of the (1/2, 0, 0) and (-1/2,0,0) peaks, while the (001) and (100) nuclear peak gains significant intensity, indicating the field induced transition to a $k = (0,0,0)$ structure. The intensity change was only observed at low-momentum transfer (lower-$Q$) reflections which confirms the magnetic origin of the transformation. It also confirms that powder sample did not reorient with applied magnetic field at this point. The Rietveld refinement was performed on the data set collected at 3 kOe and results are displayed in Figure 5b. The Figure 5b inset shows the difference between the zero field and 3 kOe powder pattern at lower $Q$. We modeled the field induced $k = (0,0,0)$ magnetic structure using the magnetic space group of $P2_1/m$ (#11.50). This symmetry puts similar constraints on the magnetic moment direction as seen for the zero-field structure, namely the Co(1) magnetic moment is parallel to $b$-axis, while the Co(2) moment can align in any crystal direction. In this model, shown in Figure 5e, the sawtooth chains align parallel to each other while Co(1) and Co(2) maintain the opposite directions within the chain. Therefore, the metamagnetic phase transition is associated with the transformation from the AFM to a ferrimagnetic state, as we observed in our magnetization measurements. The refined magnetic moments are $m_{Co(1)} = 1.8(1)$ $\mu_B$ and $m_{Co(2)} = 4.1(1)$ $\mu_B$ ( $m_a = -1.5(2)$ $\mu_B$, $m_b = 3.90(6)$ $\mu_B$, $m_c = -0.3(2)$ $\mu_B$). One may notice a slight increase in moment values compared to those at the zero field and a decrease in the canting angle for the Co(2) moments reflected in the reduction of the $a$- and $c$- moments projections. But again, the magnetization for the chain cobalt ion is close to the spin only value, while that of the vertex ion is considerably smaller.

Encouraged by our isothermal magnetization data, we carried out neutron powder diffraction up to 45 kOe applied magnetic field to look for the next field induced magnetic transitions. It is well known that the spin rotation can occur if an external magnetic field of sufficient strength is applied parallel to one of the two sublattices which is an act of minimization the energy of the system where the energy of the system could be higher if one of the AF sublattice's moments point antiparallel to the other sublattice in a sufficiently strong applied magnetic field. [59-60] Such a scenario is possible in CsCo$_2$(MoO$_4$)$_2$(OH) if the Co(1) spin starts to rotate towards the Co(2) moment direction or vice versa along the sawtooth chain direction with the continuous increase of applied magnetic field. This type of spin rotation could manifest a gradual increase of the intensity of the magnetic peaks with the applied magnetic field. Powder patterns collected form 0-45 kOe at 2 K are shown as a combined plot in Figure 6a. As mentioned in the previous discussion, we observe the suppression of (0.5,0,0) peaks with the applied magnetic field of 3 kOe. On the other hand, a significant increase of the intensity of magnetic peaks (001), (100) and (-101) associated with $\bm{k}$ = (0,0,0) structure was observed, as displayed in Figure 6a. The (001) and (-101) peaks reach a maximum intensity at around 20 kOe and then start to decrease with the further increase of the applied magnetic field. No additional magnetic peaks appeared at high fields confirming that the new phases can be described within the chemical unit cell ($\bm{k}$ = (0,0,0)) and the same magnetic space-group symmetry $P2_1/m$.

In general, it is hard to achieve a reliable magnetic moment determination for powder data under the magnetic field due to the different response from the randomly aligned grains. [60-62] At higher applied magnetic field ($H > 10$ kOe), we noticed a change in the sample texture manifested by a change of intensity of the higher Q nuclear peaks, which was addressed by refining a preferred orientation factor. We were able to follow the evolution with field of the Co(1) and Co(2) moments projected on the $b$-direction by following the change in magnetic scattering at the (001), (100) and (-1,0,1) peaks (see Figure 6a). Our measurements reveal an increase in intensity in all three aforementioned magnetic peaks as the field increased up to 30 kOe. Above this field the intensities of (001) and (-1,0,1) start to decrease while the (100) continues to gain intensity. The Rietveld refinements indicate that the Co(2) moment remains relatively unchanged in size and canting, while the Co(1) moment is slowly rotating to align parallel to the uncompensated $m_b$ or Co(2) moment component. The evolution of the magnetic structures in increasing field along the *b*-axis projection for the two Co moments is shown in Figure 6b. At approximately 35 kOe the Co(1) moment shifts its canting towards the net Co(2) spin component, along the *b*-axis, driving the second metamagnetic transition observed in isothermal magnetization data. The magnetic structure models for various fields, $H = 0 – 45$ kOe, are illustrated in Figure 6 c-g. To summarize, our neutron diffraction results indicate a first field-induced transition occurring at about 2 kOe, with the AF state defined by a wave-vector $k = (1/2,0,0)$ transforming to a ferrimagnetic state with $k = (0,0,0)$. Upon increasing the applied magnetic field the Co(1) moment rotates to form a canted FM structure. According to the magnetization data the spin canting persists up to the highest measured magnetic field of 130 kOe.

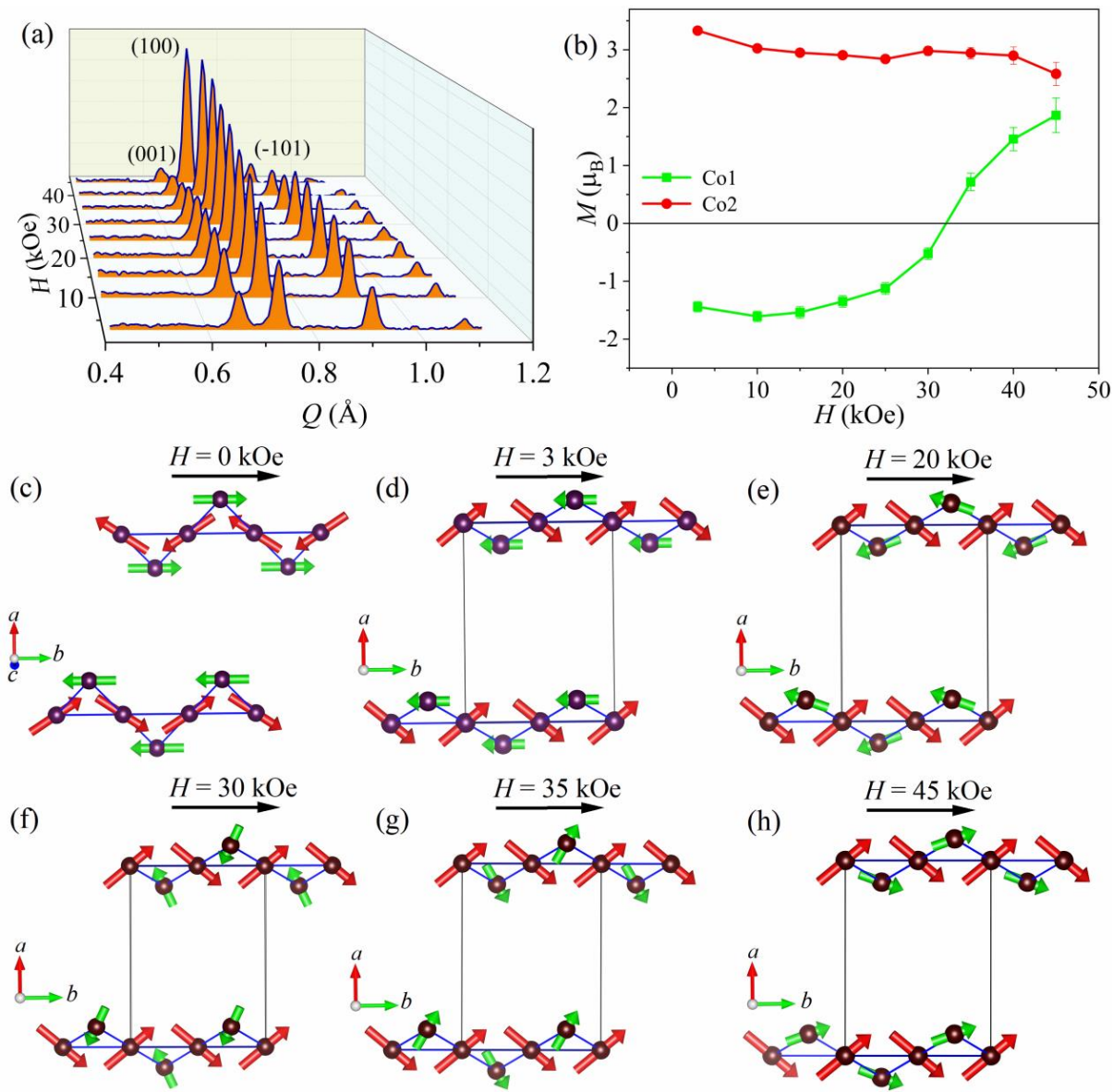

**Figure 6. Field induced states of CsCo$_2$(MoO$_4$)$_2$(OH).** (a) Neutron powder diffraction curves collected at 2 K under the applied magnetic field of 0-4.5 T. (b) Refined Co(1) and Co(2) magnetic moments projected along the *b*-axis for different applied magnetic fields. (c) Zero field magnetic structure with the ***k*** = (1/2,0,0) propagation vector. (d)-(h) The best-fit models determined from Rietveld refinements for data collected at 2 K under an applied field of 0 to 45 kOe using the ***k*** = (0,0,0) propagation vector. The magnetic structures were drawn to highlight the Co(1) and Co(2) spin orientation along the sawtooth chain direction (*b*-axis). At 3 kOe the applied magnetic field flips the next nearest neighbor sawtooth chains by rotating 180˚ to produce a ferrimagnetic structure. At 35 kOe the Co(1) spins re-orient to align with the Co(2) along the *b*-axis.

*3.5 Evaluation of the exchange parameters of $CsCo_2(MoO_4)_2(OH)$*

Consider the exchange constants $J_{bv}$ and $J_{bb}$ of a given *b*-axis chain. Whereas $J_{bb}$ acts along the spine of the chain between Co(2) atoms, $J_{bv}$ acts along the sawtooths coupling Co(2) and Co(1) atoms. Both exchange constants are positive and antiferromagnetic (AF). Assuming that the Co(1) spins $S_1$ lie in the -**b** direction and that the Co(2) spins lie at an angle q to the *b*-axis, the energy per Co(2) atom is given by

$E/N = -2J_{bv} S_1 S_2 \cos \phi + J_{bb} S_2^2 \cos 2\phi$.

Minimizing the energy E with respect to $\phi$, we find that either $\phi = \pi/2$ (Co(2) spins perpendicular to the chain) or

$\cos \phi = J_{bv} S_1 / 2 J_{bb} S_2$.

If the spins $S_1$ and $S_2$ are proportional to the Co(1) and Co(2) moments 1.3 and 3.7 $\mu_B$ with the same g constant, then the observed Co(2) canting angle of $\phi = 34°$ implies that

$J_{bv}/J_{bb} = 4.72$

or $J_{bb}/J_{bv} = 0.21$. Notice that we have specified only the component of the Co(2) spin along **b** and not the plane of the Co(2) spin. Experimentally, the Co(2) spin is observed to nearly follow one of the Co(2)–O octahedral bonds.

In a small magnetic field along **b**, each chain acts as a massive object with $N_{ch}$ Co(1)/Co(2) spin complexes that are tightly coupled to one another by the intrachain couplings $J_{bv}$ and $J_{bb}$, which are much stronger than the interchain coupling $J_{ch}$, as shown in Figure 7 (a) and (b) on the left. Based on our previous discussion, $M_{1,ch} = N_{ch} m_{1,ch}$ where $m_{1,ch} = 1.8$ $\mu_B$ is the refined moment of one Co(1) and one Co(2) spin along the *b* axis at zero field. Similarly, $M_{2,ch} = N_{ch} m_{2,ch}$ where $m_{2,ch} = 2.1$ $\mu_B$ is the refined moment of Co(1)/Co(2) just above $H_b$. Since the field $H_b = 3$ kOe is required to flip the chains from AF to ferromagnetic (FM), we obtain

$-4J_{ch} M_{1,ch}^2 = 4J_{ch} M_{2,ch}^2 - (g\mu_B)^2 M_{2,ch} H_b/2.$

Hence,

$J_{ch} = (g\mu_B)^2 H_b m_{2,ch}/8 N_{ch} (m_{1,ch}^2 + m_{2,ch}^2) = (0.0047/N_{ch})$ meV,

which scales inversely with the chain size. We can obtain rough estimates for the intrachain parameters $J_{bb}$ and $J_{bv}$ using Figure 7c. The Co(1) spins flip from -**b** to +**b** as the field increases from $H_1 = 3$ kOe to $H_2 = 45$ kOe. Since each Co(1) spin couples to two Co(2) spins through $J_{bv}$, we estimate that

$J_{bv} = g\mu_B (H_1+H_2)/4 S_2.$

Using $g = 2.8$ and $S_2 = 1.5$, we find that $J_{bv} = 0.13$ meV and using the result $J_{bb} = 0.21 J_{bv}$, we obtain $J_{bb} = 0.028$ meV. Because $H_2$ may exceed 45 kOe, these may be underestimates for $J_{bb}$ and $J_{bv}$. In any case, we have confirmed that the interchain coupling $J_{ch} = (0.0047/N_{ch})$ meV is much weaker than the intrachain couplings $J_{bb}$ and $J_{bv}$.

Note that the mean-field Néel temperature is given by

$T_N = zJ_{ch} S_{ch} (S_{ch} + 1)/3.$

Using $S_{ch} = (3/2 - 1/2) N_{ch} = N_{ch}$ for the spin of each chain at zero field and $z = 4$ for the coordination number of each chain in the *ac* plane, we find

$T_N = (4/3)\ 0.0047\ (N_{ch} + 1)$ meV $= 0.073\ (N_{ch} + 1)$ K.

Taking $T_N = 5$ K gives $N_{ch} = 68$ for the number of Co(1)/Co(2) spin complexes in each chain.

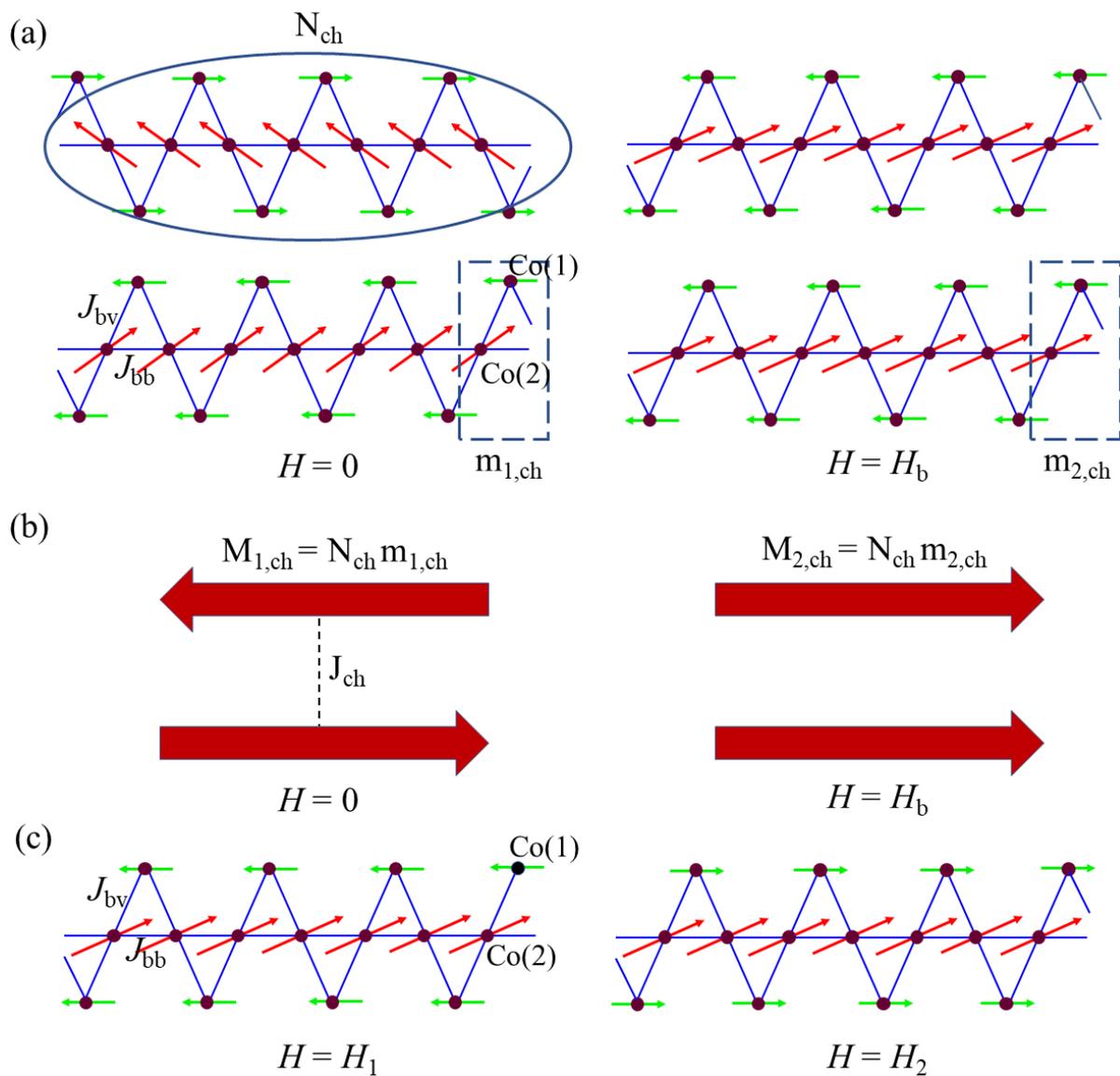

**Figure 7.** (a) Neighboring chains at zero field and at the critical field $H_b$ along **b**. (b) Those same chains drawn as large moments $M_{1,ch}$ and $M_{2,ch}$ coupled by $J_{ch}$. (c) The Co(1) spins flipping from -**b** to +**b** as the field increases from $H_1$ to $H_2$.

## 4. Conclusions

A new example of a sawtooth, or full delta, chain structure $CsCo_2(MoO_4)_2(OH)$ is reported. This general class of material does not have many members, but they all exhibit a range of complex magnetic behavior. The material can be prepared as high-quality single crystals using a high- temperature high-pressure hydrothermal method. The sawtooth structure contains a chain with one unique chain cobalt atom (Co(2)) and one unique vertex cobalt atom (Co(2)), so there are two intrachain exchanges, $J_{bb}$ and $J_{bv}$. The magnetic behavior is highly anisotropic with the chain along the $b$-axis being the easy axis, and an antiferromagnetic transition near 5 K. The material displays two successive metamagnetic transitions including a sharp one at 0.2 kOe and a broader, more gradual one around 20 kOe. This behavior was also confirmed by heat capacity measurements, which display a classic λ peak at 5.1 K that is very field sensitive and rapidly changes shape above 5 kOe.

The magnetic structure in zero field below the ordering temperature is solved as an AF phase with vectors on the vertex cobalt ions aligned parallel to the chains with both sides of the sawtooth vectors aligned in the same direction. The cobalt ions in the chain have canted vectors (34°) aligned opposite to the vertex vectors. Adjacent parallel chains have all the vectors aligned in opposite directions to create the AF phase. At the first metamagnetic transition, the magnetic moments of each chain remain the same but the moments on adjacent parallel chains flip to align in the same direction, forming a ferrimagnetic phase. Finally, as the field continues to increase ($H > 5$ kOe), the second metamagnetic transition begins to occur with Co(1) moments rotate along the $b$-axis and align in the same direction as the Co(2) creating a FM phase. The combination of neutron and magnetic data allow the calculation of the exchange parameters with the $J_{bv}$ parameter to be the largest and $J_{bb}$ to be about one fifth of that value, and the interchain exchange to be about an order of magnitude smaller still. All these behaviors are self-consistent with the magnetic properties and neutron scattering experiment.

The sawtooth chain described here has several unique properties. It is built of $Co^{2+}$ $d^7$ ions that display considerable spin-orbit coupling with magnetic entropy between that of $S = 3/2$ and $S = 1/2$, suggesting the presence of a Kramers doublet ground state. Also, the magnetic data is highly anisotropic with the direction of the chains. The magnetic phase diagram contains successive metamagnetic transitions below the ordering temperature, converting from an antiferromagnetic to ferrimagnetic structure then to a ferromagnetic structure as the field increases. Although the sawtooth structure is built of corner shared metal triangles and the original impulse might be to model the magnetic structure as a frustrated system, it is probably more useful to approach the magnetism of $CsCo_2(MoO_4)_2(OH)$ as a pseudo 1-D system with the complex magnetic behavior. Collectively, our work presents a rare example of $Co^{2+}$-based sawtooth structure which possesses complex magnetism that evolves with the temperature and the applied magnetic field. Given these observations, inelastic neutron scattering experiments, additional theoretical work and synthesis of the $Cu^{2+}$ ($S = 1/2$) analog to $CsCo_2(MoO_4)_2(OH)$ will give insight into other emergent quantum behaviors such magnetic frustration (QSL).

**Conflicts of Interest**

There are no conflicts to declare.

**Acknowledgements**:

This research used resources at the Missouri University Research Reactor (MURR). This work was supported in part by a University of Missouri Research Council Grant (Grant Number: URC-22-021). The research at the Oak Ridge National Laboratory (ORNL) is supported by the U.S.

Department of Energy (DOE), Office of Science, Basic Energy Sciences (BES), Materials Sciences and Engineering Division. This research used resources at the High Flux Isotope Reactor and Spallation Neutron Source, DOE Office of Science User Facilities operated by ORNL. The synthesis, x-ray diffraction and crystal growth at Clemson University was supported by awards from the NSF DMR – 1808371 and DMR – 2219129.